\documentclass[sigconf]{acmart}

\usepackage{balance}    %
\usepackage{color}
\usepackage{color, colortbl}
\usepackage{array}
  
\usepackage{caption}
\usepackage{subcaption}

\usepackage{multirow}

\usepackage{verbatim}  %

\usepackage{xspace}

\usepackage{graphicx}
\usepackage{xcolor}

\definecolor{Gray}{gray}{0.97}
\definecolor{MedGray}{gray}{0.9}
\definecolor{greytext}{gray}{0.5}
\definecolor{DarkGreen}{rgb}{0.0, 0.5, 0.0}
\definecolor{PFGreen}{rgb}{0.0, 0.5, 0.0}
\definecolor{lightGreen}{rgb}{0.8, 0.9, 0.8}
\definecolor{CadmiumGreen}{rgb}{0.0, 0.42, 0.24}
\definecolor{DarkKhaki}{rgb}{0.74, 0.72, 0.42}
\definecolor{DarkRed}{rgb}{0.7, 0.2, 0.2}
\definecolor{Purple}{rgb}{0.7,0.0,0.7}
\definecolor{Brown}{rgb}{0.7,0.3,0}
\definecolor{Orange}{rgb}{1, 0.5, 0.1}
\definecolor{niceblue}{rgb}{0.0, 0.2, 0.4}

\graphicspath{
{figures/} %
}

\usepackage[nameinlink,capitalise]{cleveref}
\crefname{enumi}{}{}  %
\crefrangeformat{figure}{Figures~#3#1#4--#5#2#6}

\AtBeginDocument{%
  \providecommand\BibTeX{{%
    \normalfont B\kern-0.5em{\scshape i\kern-0.25em b}\kern-0.8em\TeX}}}

\copyrightyear{2024}
\acmYear{2024}
\setcopyright{rightsretained}
\acmConference[SUI '24]{ACM Symposium on Spatial User Interaction}{October 7--8, 2024}{Trier, Germany}
\acmBooktitle{ACM Symposium on Spatial User Interaction (SUI '24), October 7--8, 2024, Trier, Germany}\acmDOI{10.1145/3677386.3682094}
\acmISBN{979-8-4007-1088-9/24/10}

\begin{document}

\title{Social VR for Professional Networking: A Spatial Perspective}

\author{Victoria Chang}
\email{vchang90@umd.edu}
\affiliation{%
  \institution{University of Maryland}
  \city{College Park}
  \state{Maryland}
  \country{USA}
}
\authornotemark[1]

\author{Ge Gao}
\email{gegao@umd.edu}
\affiliation{%
  \institution{University of Maryland}
  \city{College Park}
  \state{Maryland}
  \country{USA}
}
\authornotemark[1]

\author{Huaishu Peng}
\email{huaishu@umd.edu}
\affiliation{%
  \institution{University of Maryland}
  \city{College Park}
  \state{Maryland}
  \country{USA}
}

\authornote{All authors contributed equally to this work and are listed in alphabetical order.}

\renewcommand{\shortauthors}{Chang et al.}

\begin{abstract}
One essential function of professional events, such as industry trade shows and academic conferences, is to foster and extend a person's connections to others within the community of their interest. In this paper, we delve into the emerging practice transitioning these events from physical venues to social VR as a new medium. Specifically, we ask: how does the spatial design in social VR affect the attendee’s networking behaviors and experiences at these events? To answer this question, we conducted in-situ observations and in-depth interviews with 13 participants. Each of them had attended or hosted at least one real-world professional event taking place in social VR. 
We identified four elements of VR spatial design that shaped social interactions at these events: area size, which influenced a person's perceived likelihood of encountering others; pathways connecting areas, which guided their planning of the next activity to perform; magnets in areas, which facilitated spontaneous gatherings among people; and conventionality, which affected the assessment of a person's behavior appropriateness. 
Some of these elements were interpreted differently depending on the role of the participant, i.e., event hosts vs. attendees. We concluded this paper with multiple design implications derived from our findings.
\end{abstract}

\ccsdesc[500]{Human-centered computing ~Human computer interaction (HCI)}
\ccsdesc[300]{Human-centered computing ~Empirical studies in HCI}

\keywords{Social VR, Professional event, Social interaction, Spatial design}

\maketitle

\section{Introduction}
The democratization of virtual reality (VR) technology has opened new avenues for individuals to connect with one another virtually and immersively. Leveraging commercially available social VR platforms, such as VRChat, people today can gather for various social events without geographical barriers ~\cite{sykownik2021most, mcveigh2018s, perry2015virtual}. While many of these events cater to casual entertainment, there has been a noticeable increase in the number of professional networking events hosted in virtual spaces, including virtual trade shows and academic conferences ~\cite{moreira2022toward, mulders2021academic, erickson2011synchronous}.

Building and strengthening social connections in the above scenarios usually requires people to follow certain norms that fit into the atmosphere of the professional event ~\cite{edelheim2018conferences, levine2015networking}. To this end, much previous work has emphasized VR's benefit in terms of granting a person the full freedom to tailor their appearance. It has been reported through a big body of studies that, by choosing and updating their avatars, people can craft the identities and images they wish to project and, thereby, shape the way others may interact with them (e.g., ~\cite{heidicker2017influence, freeman2021body}). In contrast, little research has closely examined the connection between social interactions at a professional event and VR's capability to (re)configure the spatial environment in which these social interactions take place. 

Since the late 1960s, environmental psychologists have confirmed that the spatial design of a physical venue, e.g., a conference hall's layout or furniture arrangement, can provide subtle yet noticeable guidance for social interactions at the venue~\cite{barker1968ecological, meagher2020ecologizing}. For instance, during coffee breaks at offline professional events, participants often circulate around dining tables to engage in conversations. The array of tables in an open hallway acts as tacit social cue, encouraging participants to gather around and seek networking opportunities with those nearby.

As professional events go increasingly virtual today, it poses both opportunities and challenges for VR spatial design that can best facilitate people's networking activities. On the one hand, VR makes it possible for all users, especially those hosting the event, to customize the venue's spatial features with little constraint; on the other hand, given the default discrepancies between the physical and digital worlds, people may not fully adopt social conventions established in the former setting to guide their interactions in the latter. It is, therefore, critical to ask: What is the relationship between the spatial design of a social VR environment and professional networking activities taking place in that environment? Which specific aspects of the spatial design matter and how?

The current paper provides in-depth qualitative insights into the interplay between spatial design and professional networking in a social VR context. Specifically, we performed in-situ observations and interviews with 13 participants. Each participant had attended or hosted one or more professional events arranged via a social VR platform, covering a total of 20 professional events. These methodological choices allow us to delve into people’s in-situ practices and experiences at real events hosted in VR, complementing prior work that examines the effects of VR spatial design via controlled experiments or user studies of in-house systems (e.g.,~\cite{iachini2014body, choudhary2021revisiting, bonsch2018social,kalisperis2006evaluating}).

Our data pointed to four aspects of VR spatial design that contextualize people’s social interactions at professional events: the size of an area, shaping person’s belief in their chance to bump into others; the pathway between areas, guiding a person’s planning of their next activities; the magnet placed in an area, enabling a person’s self-initiation of temporary gatherings; and the conventionality of the scene, assisting the assessment of a person's behavior appropriateness. 

Across these findings, we demonstrated two fashions of critical misalignments disrupting people’s sensemaking of the social meanings attached to spatial cues. Some of these misalignments existed between the objective spatial status of a virtual venue and the event participant’s spatial perception while navigating the venue (e.g., people perceived others to be far away when the virtual geometric distance was not long). Others revealed the misalignment between what event hosts intended to achieve through spatial arrangement ahead of an event and how attendees interacted with the space at the event (e.g., hosts designed virtual pathways directing people to an activity area, but attendees leveraged teleporting to skip them). 

With these findings in mind, we discuss potential design solutions to facilitate the synchronous interpretations, as well as negotiations, of spatial-as-social-cues between the VR system and users taking up various roles at the event. Our work contributes to the growing body of social VR and spatial user interface research by adopting a human-centered approach.

\section{Related Work}

\subsection{Professional Networking and Spatial Design in Office Settings}
Professional events are public events that allow individuals to build and strengthen their connection to a given field. One essential dimension of this connection is networking, or the establishment of contact among people.

Despite its significance, networking can be challenging~\cite{owens2008you, van2014gender, mitchell2016should}. Previous studies have found that attendees at business mixers sometimes lack clues about when and where to engage with other individuals of interest~\cite{ingram2007people}. During social hours at professional conventions, it is common to observe people conversing primarily with the person directly beside them rather than branching out to others~\cite{marroun2016professional}. These findings underscore the need to assist individuals in navigating the social aspect of professional events, such as identifying opportunities to establish rapport, capturing others’ attention for interactions, and understanding the norm of behavior.

The spatial design of a physical venue can provide tacit yet discernible guidance for networking behaviors within the venue. This notion was first introduced by environmental psychologists in the late 1960s (e.g., \cite{barker1968ecological}) and has been repeatedly verified by empirical research over subsequent decades (see ~\cite{meagher2020ecologizing} for a review). For instance, Forgas and Brown’s lab experiment indicated that, compared to open public spaces, indoor settings would increase a person’s sensitivity to interpersonal cues disclosed by others \cite{forgas1977environmental}. Pfeffer performed fieldwork to identify stimuli and inhibitors of social interactions in office buildings. He found that the placement of partitions often constrained a person’s choice regarding whom to interact with and when \cite{davis1984influence}. Evans and McCoy investigated the link between spatial design and psychological stress. They found that neutral territories in an environment could serve as buffer zones, alleviating a person’s mental burden before engaging with others in more designated areas \cite{evans1998buildings}. Ju and Takayama explored the social implications of objects, such as gesturing doors, installed in a space. Their research revealed that individuals tended to interpret certain movements of the object as signals inviting them to join social activities in the space \cite{ju2009approachability}. Their research revealed that individuals tended to interpret certain movements of the object as signals inviting them to join social activities in the space (e.g., \cite{ching2023architecture, mandeno2022six, harris2014some}). 

Other scholars and practitioners have applied the insights mentioned above to the specific context of event planning. In particular, Ching discussed the social function of circulation spaces, such as hallways and corridors, within a large venue hosting professional events. Given its role in bridging different segments of a whole, circulation space carries the potential to indicate and foster opportunities for spontaneous encounters~\cite{ching2023architecture}. Mandeno and Baxter explored the structural complexity of a space and its influence on networking behaviors. According to them, an ideal space for social interaction should adhere to the principle of minimizing distractions. Placing too much furniture in the venue can divert people from their primary goal of connecting with one another~\cite{mandeno2022six}. Marroun reflected on the interactions that often occurred during cocktail hours and sit-down dinners at professional events. From there, they emphasized the importance of seating arrangements and flexibility in shaping physical proximity as well as interpersonal distance among event attendees~\cite{marroun2016professional}.

\subsection{Social VR as a New Medium for Professional Networking}
Today, the advancement of social VR technology is shifting the venues, as well as the professional events hosted in those venues, from the physical to virtual environments (e.g., \cite{curtis1994muds, mcveigh2018s, maloney2020talking, sykownik2022something, freeman2021body, freeman2022working, saffo2021remote, orlosky2021telelife}). ``Going virtual'' offers event attendees the convenience of interaction without physical boundaries while also raising the unknown of to what extent the social functions situated in the real world are transferred to the digital counterpart.

Among the growing body of literature on social VR, some suggest an equivalence between virtual and physical spaces in influencing people’s networking behavior. For instance, recent studies from McVeigh-Schultz and colleagues showed that social VR users often drew upon the mental model established from a physical space to infer the social norms and possibilities of acting within its virtual counterpart \cite{mcveigh2019shaping, osborne2023being}. When a virtual space designed for group events employed the skeuomorphic design of a physical auditorium, those entering the events would typically remain quiet and listen to the person standing in the center \cite{mcveigh2019shaping}. In a similar vein, Bonfert et al. analyzed survey responses from 32 individuals who utilized various social VR platforms (e.g., AltspaceVR, Engage) for professional gatherings during the COVID pandemic. Participants reported that they identified the opportunity to join an ongoing conversation by carefully observing the spatial relationship among interlocutors. Such a tactic mimics how similar decisions were made at events in physical venues~\cite{bonfert2023seeing}. 

Meanwhile, virtual spaces also possess attributes that distinguish them from the real world. For example, one attribute heavily discussed in HCI literature is the lack of convincing, synchronized multisensory feedback for spatial perception~\cite{jung2021use, jung2020impact}. Individuals in a virtual space may have to interpret the social meaning of spatial information based on incongruent clues. As one example, Williamson et al. hosted an academic event in their customized space via Mozilla Hubs. Although many attendees intended to manage the interpersonal dynamics between themselves and others properly, they reported challenges in determining the actual location of another avatar when it was not in their direct line of sight~\cite{williamson2021proxemics}. Similar observations were also noted in the report from a large-scale virtual event within the IEEE VR community~\cite{moreira2022toward}.

The inherent malleability of virtual environments also distinguishes them from offline environments. Unlike physical spaces, VR environment grants users the liberty to modify or even reconstruct a given space~\cite{osborne2023being, freeman2022working, sermarini2022bim}. The authority to assign social meanings to a particular spatial design, which was traditionally under the control of event hosts, is now possible to be distributed among everyone attending the event, at least to some extent.

The literature reviewed above, along with our reflections, calls for empirical research at the intersection of spatial design, networking behavior, and social VR.  In the remainder of this paper, we present our recent endeavor to embark on this research. Our work contributes qualitative insights that respond to the following research question (RQ): 

\textit{How does spatial design shape people's networking behaviors and experiences at professional events that are hosted in social VR?}

\section{Method}
We performed qualitative research with 13 participants and at 20 professional events. Throughout the data collection phase, the first author of this paper observed each participant during one or more events that the participant had planned to attend before enrolling in our research. Observations were carried out in the format of natural shadowing. Specifically, the researcher maintained a proper distance from the participant while observing the participant's behaviors at the event within the virtual space. She also avoided initiating personal interactions with the participant, unless the participant chose to approach her for casual conversations. 

Immediately after each observation, the researcher interviewed the participant to understand their first-person perspectives considering the spatial setup of the social VR in use as well as its influences on their networking experiences at the event. 

All observational and interview data were then analyzed following the process suggested by the thematic analysis method~\cite{braun2006using}. The research received approval from the Institutional Review Board (IRB) of the author's institution. In the following, we detail our recruiting strategy, the observation and interview protocols, and the data analysis process. 

\subsection{Recruiting Strategy}

Our recruitment took place through the dissemination of digital flyers via two primary channels: internal mailing lists within the researcher's institutions (e.g., those used for announcing VR-relevant research talks and events to all self-registered audience members) and public online interest groups (e.g., the VR enthusiasts group on Facebook and Discord servers). Eligible participants had to be at least 18 years old and comfortable communicating with the researcher in English. 

Besides, we intentionally included a) individuals who took various roles (i.e., hosts or attendees) at their self-identified events and b) events that spanned multiple social VR platforms (i.e., AltSpaceVR~\footnote{AltSpaceVR: https://web.archive.org/web/20190723190411/https://altvr.com/new-avatars/} , VRChat~\footnote{VRChat: https://hello.vrchat.com/}, or Venu~\footnote{Venu: https://www.venu3d.com/}) and different professional communities (i.e., art and design, tech development, and academia). 

Table~\ref{tab:recruitment} details the background information of all the 13 participants, including their ID, role at the event, platform used to join the event, professional community, number of observed events in their professional community, age, gender, overall experience of attending social VR events, and overall experience of using VR technology.

\begin{table*}[]
\caption{Background information of all participants.}
\resizebox{\textwidth}{!}{%
\setlength\extrarowheight{10pt}
\begin{tabular}{lllllllll}
\hline
\multicolumn{1}{c}{\textbf{ID}} & \multicolumn{1}{c}{\textbf{Role}} & \multicolumn{1}{c}{\textbf{Platform}} & \multicolumn{1}{c}{\textbf{Community}} & \multicolumn{1}{c}{\textbf{\# of Observed Events}} & \multicolumn{1}{c}{\textbf{Age}} & \multicolumn{1}{c}{\textbf{Gender}} & \multicolumn{1}{c}{\textbf{Exp. of Attending SVR Events}} & \multicolumn{1}{c}{\textbf{Exp. of Using VR Technology}} \\ \hline
P1                              & Attendee                          & VRChat, AltspaceVR                    & Art and design                         & 2                                                 & 25-34                            & Female                               & Several times a year                                      & Several times a year                                     \\
P2                              & Attendee                          & VRChat, Venu                          & Tech development                       & 3                                                 & 25-34                            & Male                                 & Several times a year                                      & Several times a month                                    \\
P3                              & Attendee                          & AltspaceVR                            & Academia                               & 1                                                 & 18-24                            & Female                               & several times a year                                      & Several times a month                                    \\
P4                              & Attendee                          & VRChat                                & Art and design                         & 2                                                 & 25-34                            & Female                               & Several times a month                                     & Several times a month                                    \\
P5                              & Attendee                          & AltspaceVR                            & Academia                               & 1                                                 & 35-44                            & Male                                 & Several times a year                                      & Several times a year                                     \\
P6                              & Attendee                          & AltspaceVR                            & Academia                               & 1                                                 & 18-24                            & Male                                 & Several times a year                                      & Several times a year                                     \\
P7                              & Attendee                          & Venu                                  & Tech development                       & 1                                                 & 35-44                            & Male                                 & Several times a month                                     & Several times a week                                     \\
P8                              & Attendee                          & Venu                                  & Tech development                       & 1                                                 & 25-34                            & Male                                 & Several times a month                                     & Several times a week                                     \\
P9                              & Host                              & VRChat, AltespaceVR                   & Art and design                         & 2                                                 & 18-24                            & Male                                 & Several times a week                                      & Several times a week                                     \\
P10                             & Host                              & Venu                                  & Tech development                       & 1                                                 & 25-34                            & Male                                 & Several times a week                                      & Daily                                                    \\
P11                             & Host                              & Venu                                  & Tech development                       & 1                                                 & 25-34                            & Male                                 & Daily                                                     & Daily                                                    \\
P12                             & Host                              & VRChat                                & Art and design                         & 2                                                 & 35-44                            & Male                                 & Several times a week                                      & Several times a week                                     \\
P13                             & Host                              & AltspaceVR                            & Academia                               & 2                                                 & 45-54                            & Female                               & Several times a week                                      & Several times a week                                     \\ \hline
\end{tabular}%
}

\vspace{2ex}
\raggedright \footnotesize \textbf{*Note:} In the rest of this paper, we use the combination of [ID, role, platform], as specified in the first three columns of Table~\ref{tab:recruitment}, to refer to a participant when presenting interview responses quoted from them.
\label{tab:recruitment}
\end{table*}

\subsection{Observation and Interview Protocols}
We generated two semi-structured protocols to guide our data collection with each participant.

\textit{Observation protocol}. With consent, the researcher attended the event specified by that person and observed their behaviors. The duration of each event ranged from 1 to 1.5 hours. The total number of attendees at a given event was between 10 and 50. The researcher obtained written or pictorial notes to document a) the spatial design of the social VR in use, b) any of the participant's behaviors that she would like to discuss in post-event interviews, and c) her own experience at this event as an observer. Participants were always invited to review these observation notes after the events. Should the participant express discomfort with any of the noted information, that information must be removed from the record and the subsequent data analysis.  

\textit{Interview protocol}. We interviewed each participant within 24 hours following the event they had attended. All interviews took place via Zoom. Each lasted between 30 minutes and 1 hour. Questions discussed during the interview centered around three topics: a) the participant's overall experience at the given event, b) their takeaways of leveraging any spatial clues to navigate the interaction with others, and c) their reflections of any moments or behaviors highlighted in the researcher's observation notes. We audio-taped and transcribed these interviews for analytical purposes.

\subsection{Data Analysis}
We performed thematic analysis to make sense of our data. 

At the beginning, all members of our research team read the observation notes and interview transcripts thoroughly, familiarizing ourselves with the data. The first and second authors of this paper then independently reviewed distinct subsets of the data. They generated an initial set of codes that captured the semantic content of the data as well as the latent notions it revealed. These codes described the virtual context of each observed interaction at a given event, the social behavior performed by participants in that context, the perceived success or obstacles of the interaction, and the comparison between professional networking events in social VR and physical reality.

After that, the research team held a series of group coding sessions, collaboratively applying the initial codes across the entire dataset. At these sessions, we verified, refined, and expanded the developed codes through extensive discussions. We repeated these steps until reaching a point of saturation, where no new codes emerged and all redundant codes were systematically removed.

\section{Findings}
Overall, participants in our research all explicitly stated that they opted to network with their professional community via social VR \textit{“for its spatial features.”} Many expanded on this by contrasting social VR with two other media: video conferencing and 2D virtual environment. While video conferencing platforms, like Zoom, are adept at facilitating structured conversations, they often lead to complaints about a \textit{“loss of autonomy.”} 

In particular, our participants characterized their networking experience via Zoom and similar platforms as \textit{“feeling trapped in gridded squares,”} \textit{“having minimal choices in staying further away or approaching a specific person,”} and \textit{“should always be prepared for the next activity set by the host.”} The absence of spatial information was emphasized as the very barrier preventing the transfer of soft skills between real-life and virtual events. 

Some participants recounted their use of Gather.town, a virtual environment that offers spatial information of virtual venues but in a 2D format. To them, the spatial design of Gather.town provides its users with more control in navigating the geometrical layout of the event, rather than the nuances of social interactions. 

Social VR stands out from other media because it enables our participants to discover networking opportunities as those opportunities \textit{“naturally unfold during the exploration of space.”} The bulk of our data revealed four specific elements of spatial design in social VR that affect participants’ professional networking behaviors at the events they attended. In the following, we provide four findings from both the event hosts and the event attendees' perspectives.

\subsection{Size of an Area, Chance of Bumping into Others}\label{resize}

\subsubsection{``This venue has been re-sized a few times for different crowds.''}
Like in a real physical space, the size of an area within the virtual venue can shape participants’ intuitive assumptions about their likelihood of encountering others. A virtual area that maintains a proper perceived interpersonal distance may facilitate networking, whereas a large, open area may appear too empty for participants to engage in networking effectively. 

From our observation notes, we documented numerous cases in which attendees wandered around large lobbies or gallery halls at their event venue, then exited very swiftly without engaging with others. Interview responses from P2 and P6 provided insights into the rationale behind such behavior:

\begin{quote}
    \textit{``As a game developer, I have attended many events in my community in VR. One thing I have learned is that if you ‘walk into’ some large rooms, there is no need to stay there long. The room would quickly feel very spread out and inactive. There might be clusters of people talking, but they usually seem distant from you.  It just didn't feel very inviting. At the particular moment you asked about, I was about to leave because I felt it was difficult to approach someone for a chat.''} [P2, Attendee, Venu]
\end{quote}

\begin{quote}
    \textit{``Large rooms usually felt empty. Or, there might have been a few people clustered and talking, but they may just spread out before you could move closer. It is just harder [to initiate an interaction]. The vibe feels like people might not want to talk, or I might not be able to join the small group they have already formed.''} [P6, Attendee, AltSpaceVR]
\end{quote}

While ample space in a real physical setting might also deter participants from interacting with each other, there is no means to ``quit'' the event promptly with just a click of a button. Thus, the size of a virtual area is arguably more critical than that of a real physical event in keeping participants engaged within the event.

Many event hosts in our observation were well aware of the interplay of the virtual area size and the likelihood of effective networking. To better facilitate networking opportunities, they would leverage the customization features in VR and adjust the size of each virtual area \textit{“to its best optimal”}. 
For example, hosts reported that, prior to an event, they would collect information to estimate the expected number of attendees for different activities at the event. 
This data enabled them to \textit{“resize all the rooms appropriately in between events”}—hoping that the room size would neither be \textit{``too large, causing people to spread out too much,''} nor \textit{``too small that people are crammed together.''} As P11 and P12 specified:

\begin{quote}
    \textit{``Many VR applications allow people to build and modify the [virtual] world for event planning. As hosts, we can change the size of the area without having to bring in real, raw materials or tear down a real wall, or wait for booking centers to confirm for a convention center. We can customize our own, and choose which fits best for the number of people attending each virtual talk, poster session, or coffee break.''} [P11, Host, Venu]
\end{quote}

\begin{quote}
    \textit{``We have a few types of activities we've run repeatedly across events, like panelist talks or purely networking hours. We notice that there is a general size of the crowd for each activity. We can also get a feel for how many people are coming through email sign-ups or Meetup sign-ups. This helps us determine if a room should be enlarged or sized down for the upcoming event. But it won’t be precise because things can change from one event to another.''} [P12, Host, VRChat]
\end{quote}

It should be noted that, while fine-tuning the size of virtual areas represents a potential solution in VR, some event hosts have highlighted its limitations. First, determining the appropriate size of a space is largely based on experience. More experienced event hosts may be able to estimate an appropriate space size, but for many, the specific dimensions of a virtual space and how many attendees it can accommodate are not clear and require trial-and-error. Second, although resizing the space of an event is possible, it is an ``always delayed'' measure. Editing virtual space takes time, and it is impractical to alter the size of the space while it is occupied. Consequently, if the number of attendees significantly deviates from initial expectations, hosts lack effective means to resize an virtual area in real-time.

\subsubsection{``They sound close but look far away.''}\label{audio}
Comparing our observation notes with participants’ reports from the same events, we noticed that the perceived size of a virtual room may not remain constant as the physical size of a real-world room would. Rather, it can vary according to specific cues that each individual adopts to shape their own perception.  As one example, P1 reflected on her experience of attending a virtual exhibition organized within a community of 3D artists. She described part of that event as follows:
\begin{quote}
    \textit{``It was very chaotic and disorienting, as I could hear everyone’s conversations around me. It was a bit frustrating when you felt so many voices were approaching you equally. You  just didn't know where to move to.''} [P1, Attendee, AltSpaceVR]
\end{quote}

\begin{figure}[ht]
  
  \includegraphics[width=\columnwidth]{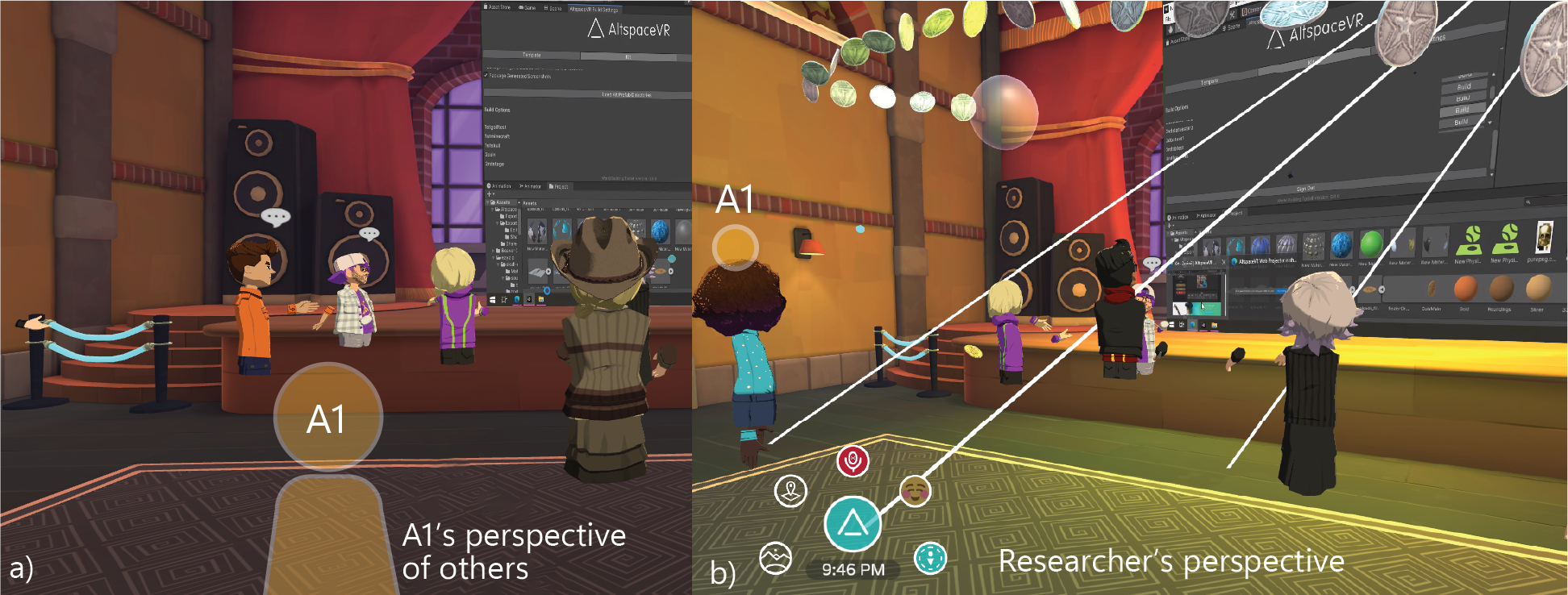}
  \caption{The left image shows P1's perspective of the virtual room containing herself and five others. On the right is the view according to the researcher’s perspective, where everyone in this room kept proper distance from each other. }
  \Description{The left image shows P1's perspective of the virtual room containing herself and five others. On the right is the view according to the researcher’s perspective, where everyone in this room kept proper distance from each other. }
  \label{fig:audio}
\end{figure}

Notably, the exact event mentioned by P1 did not appear crowded according to the researcher’s experience observing it (Figure~\ref{fig:audio}). The area where P1 spawned had ample space, with five individuals maintaining appropriate distances from one another. However, because the spatial audio at this virtual event was not properly set up to align with visual cues, our participant heard voices coming from people at various distances, all at volumes that were indistinguishable. This dissonance between spatial information in visual and audio formats led to the feeling that she was closely surrounded by others. Similar experiences were also reported by several other participants, albeit infrequently.

\subsection{Pathways Connecting Areas, Planning of the Activity to Join}\label{pathways}

\subsubsection{``It gives time to prep.''}
As one of the main goals of professional networking events is to facilitate meet-ups and conversations among attendees, event hosts often structure the VR space into different rooms and areas for people to meet. While virtual rooms are certainly where attendees meet each other, our observation and interview data both underscore that the pathways—virtual spaces that bridge different virtual areas—are also key to networking events and serve multifaceted social functions.

Specifically, event hosts in our sample exhibited a high awareness of a pathway’s potential in \textit{“enabling natural encounters”} among individuals during professional networking events. 
We observed at multiple events that attendees would occasionally cross paths while traversing virtual bridges or tunnels to move between activity areas. During these moments, many would pause to engage in small talk with each other. As we discussed such observations with participants who were hosts, they shared the sentiment that the interaction \textit{“is part of the [host’s] plan.”} For example, P12 elaborated as follows:
\begin{quote}
    \textit{``As event hosts, we put roads and bridges there for a reason. We planned our virtual space in that way because people will follow those paths, and that's where they will meet. If you see someone you want to talk to, just make a pitch to them while making your way over to the other side of the bridge.''} [P12, Host, VRChat]
\end{quote}

\begin{figure}[ht]
  
  \includegraphics[width=\columnwidth]{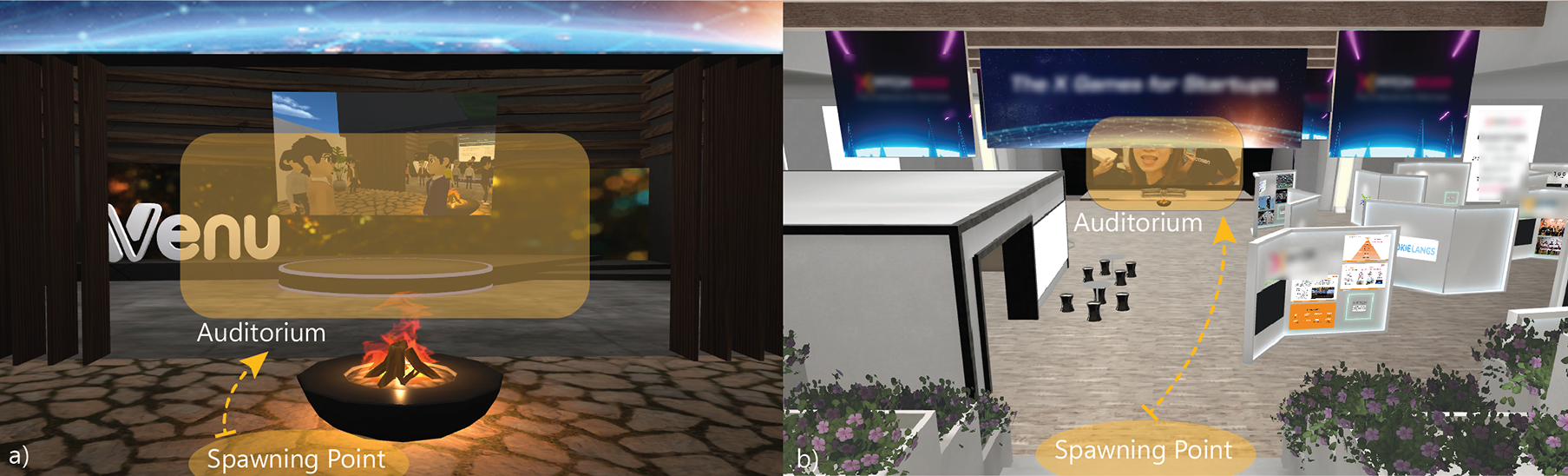}
  \caption{The left image, provided by the participant, shows one version of the space's layout where attendee's spawn point was adjacent to the auditorium. The right image shows an alternative version that used an extended pathway to connect various activity areas.}
  \Description{The left image, provided by the participant, shows the initial layout where attendee's spawn point was adjacent to the auditorium. The right image shows the revised layout with an extended pathway.}
  \label{fig:resizepassway}
\end{figure}

In addition, many hosts commented on the role of pathways in helping attendees decide and be mentally prepared for the next activity to join. P11’s interview offers a representative example in this context. 
As illustrated in Figure~\ref{fig:resizepassway}, P11 has re-designed the virtual venue to host professional events within his community over multiple times. 
Initially, the layout positioned the attendee’s spawn point adjacent to the auditorium, as he believed a pathway between the two areas was \textit{“not necessary.”} 
However, after observing attendee behavior during several events, he generated an alternative spatial design. The second version of the space's layout employs an extended pathway to connect various activity areas within the venue. 
By doing so, attendees \textit{“can now see into an area and adjust their anticipation about what is already happening or about to happen in that area.”}

\begin{figure}[ht]
  
  \includegraphics[width=\columnwidth]{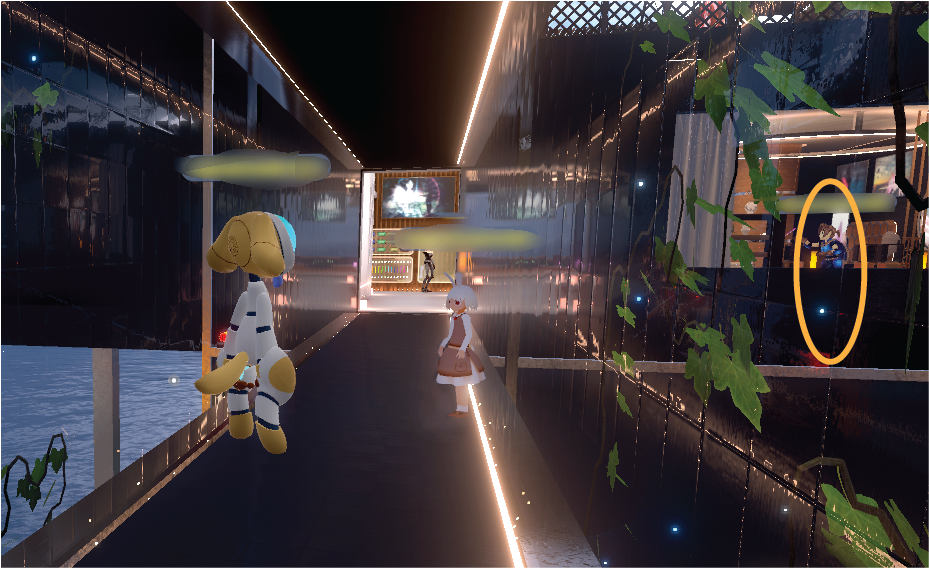}
  \caption{
P4 traversed an aisle en route to the other side of the event venue, where P4 encountered someone else and exchanged greetings. Subsequently, P4 noticed another attendee (highlighted in yellow) through the glass walls of the aisle.}
  \Description{
P4 traversed an aisle en route to the other side of the event venue, where P4 encountered someone else and exchanged greetings. Subsequently, P4 noticed another attendee (highlighted in yellow) through the glass walls of the aisle.}
  \label{fig:pathway}
\end{figure}

Our observation of P4 provides another vivid example, this time from the attendee’s perspective (Figure~\ref{fig:pathway}). 
During this observation, we noticed that P4 traversed an aisle en route to another side of the event venue. 
Two notable social encounters occurred during this brief moment. 
First, P4 encountered another individual standing in the aisle. Without being prompted, both parties oriented their avatars toward each other and exchanged greetings. 
Subsequently, P4 approached the shadowing researcher and said, \textit{“you see the blue person in that room? It seems something fun is going on there. I will go to check it out!”} Evidently, the extended aisle, coupled with the surrounding glass walls, afforded this participant the opportunity to strategize and prepare for the next interactions with others. From a distance, the researcher did observe that P4 interacted with the individual in blue upon entering the room, although the specifics of that interaction were not captured in our data record.

\subsubsection{``But teleport is faster!''} \label{teleport}
Pathways in the venues we observed were often designed to facilitate networking among attendees. 
Nevertheless, it is worth noting that not all individuals utilized these pathways to move between areas. 

We observed a handful of instances in which attendees opted to teleport themselves from one location to another during professional events in social VR. 
The following quotations from P6 and P8 detailed our participants’ reasons to choose teleportation:
\begin{quote}
    \textit{``At the beginning, I explored the virtual event by walking because I didn't know where things were. But as I became more familiar with the layout, I started using teleportation. It's a much faster way to get around the entire space and to see the whole event. It helps make sure I won’t miss out on anything or overlook people I could talk to.''} [P6, Attendee, AltSpaceVR]
\end{quote}

\begin{quote}
    \textit{``Walking in VR takes time. By teleporting, it doesn't matter how big this virtual world is. I can just teleport to the auditorium or a poster booth as long as my pointer can reach it, which it did. When I am in a hurry in the real world, I can walk or run. But sometimes in VR, you're limited to only two speeds, and you might be late to the next activity at the other side of the hallway. I'm less anxious about getting somewhere on time using teleportation.''} [P8, Attendee, Venu]
\end{quote}

Meanwhile, because teleportation does not allow for precise route planning or landing spots on the VR platforms where the events were hosted, participants sometimes find themselves \textit{“suddenly appear out of nowhere.”} 
As P1 indicated:
\begin{quote}
    \textit{``A downside of teleportation is that you have no bearing of what’s around you. I have experienced some really awkward moments when I just appeared in the middle of a group. People were already having a conversation somewhere in the (virtual) room, and I just showed up out of nowhere. It was embarrassing for everyone.''} [P1, Attendee, AltSpaceVR]
\end{quote}

\subsection{Magnets in an Area, Creation of Temporary Gathering Spots}\label{magnet}

\subsubsection{``Everyone just moves over to the bonfire.''}
Walls and rooms function as mechanisms for segmenting large spaces into designated small areas and for directing people toward specific social activities within those areas. Participants in social VR events continue to rely on these conventional and usually permanent clues to navigate their meet-ups with others. 

\begin{figure}[ht]
  
  \includegraphics[width=\columnwidth]{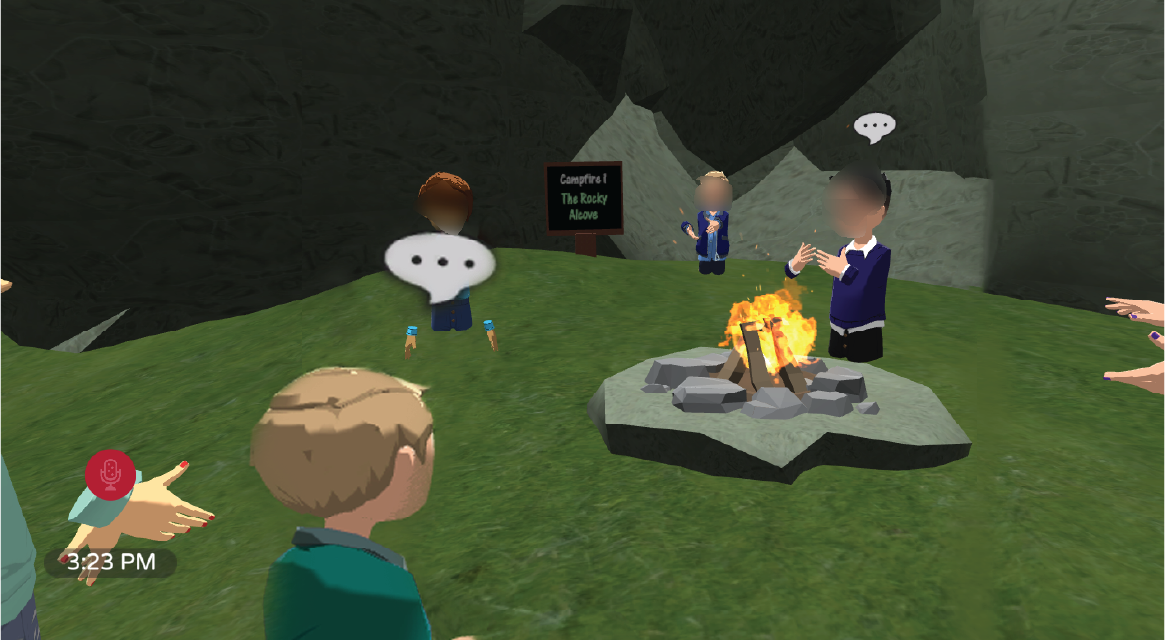}
  \caption{The bonfire caught the attendees' attention and led them to naturally gather in small groups.}
  \Description{The flame in the bonfire stood out from the rest of the environment, caught the attendees' attention, and led them to naturally group around the bonfire.}
  \label{fig:bonfire}
\end{figure}

Nevertheless, we found that VR participants have developed practices of introducing self-crafted artifacts, which we term “magnets,” to initiate social networking opportunities in a temporary and flexible manner. Figure~\ref{fig:bonfire} illustrates a case where one of the event hosts from our sample, P9, strategically arranged multiple virtual bonfires to signify that \textit{“this open area is good for people to stand together and have small talks.”} As we witnessed from the subsequent moments, the flame in the bonfire successfully garnered the attendees' attention. Many naturally gravitated toward the bonfire, upon which the host initiated greetings and facilitated interactions, all the way until the attendees moved on to their next activities in other virtual areas.

In interviews with other participants, we heard people talking about similar cases but from the perspective of the attendees. For instance, P5 and P6 each shared their experiences of forming small groups around virtual objects, specifically a bonfire and an eye scanner, which served as magnets: 

\begin{quote}
    \textit{``I spotted a few other people gathered around a bonfire hanging out. I know we were in VR but I still felt it was natural to move toward the bonfire. So, I headed over there. The atmosphere was nice, and I felt welcome there immediately.''} [P6, Attendee, AltSpaceVR]
\end{quote}

\begin{quote}
    \textit{``There's always a question of how you might organize a crowd during VR events. What happened at today’s event was interesting. They placed an eye scanner there, like you would see in a movie where you scan your retina and the door opens into somewhere. People actually lined their avatars up to do it, and so did I. It's fascinating because there was nothing actually scanning. It was not real, but people just wanted to do it. We lined up and started to talk to each other spontaneously.''} [P5,  Attendee, AltSpaceVR]
\end{quote}

\subsubsection{``I made a flower to attract people.''}
Interestingly, our data showed that the creation of magnets was not solely of interest to hosts. Several attendees also expressed a willingness to introduce their own magnets into the event space. 

\begin{figure}[ht]
  
  \includegraphics[width=\columnwidth]{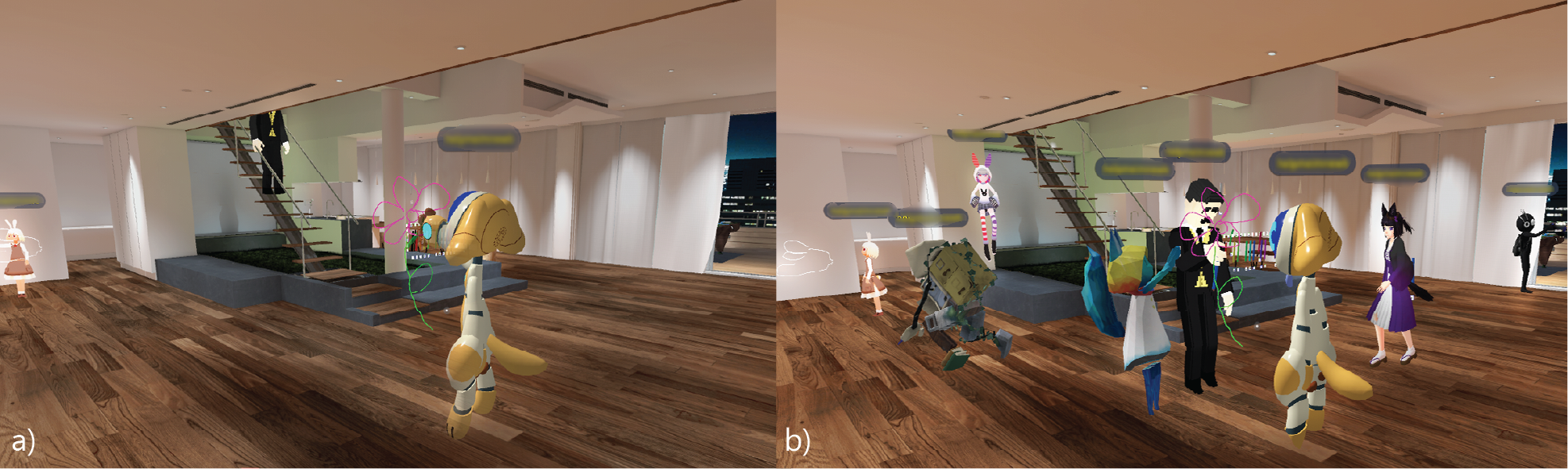}
  \caption{The left image shows that P4 was observed drawing the virtual flower. The right image shows that the flower drawing attracted others to P4's spot, where conversations were initiated.}
  \Description{The left image shows the moment when P4 was observed drawing the virtual flower. The right image shows the subsequent moment when the flower drawing attracted others to P4's spot for conversations.}
  \label{fig:flower}
\end{figure}

P4 was the only attendee in our sample who actualized this intention. In P4's specific case, the magnet P4 created was a virtual flower. It served not only as a spatial signal to attract others to the participant's spot, but also as an icebreaker to initiate conversations (Figure~\ref{fig:flower}). During the post-event interview, P4 elaborated that:

\begin{quote}
    \textit{``At all the events I’ve attended in VR, I was never allowed to create things like sculptures or tables. But I found we could ‘draw an object’ in mid-air. You can [do the drawing to] make connections with people because they want to see what you've created. That gives you an opportunity to interact with them.''} [P4, Attendee, VRChat]
\end{quote}

\subsection{Conventionality of the Scene, Self-Assessment of Behavioral Propriety}\label{conventionality}

\subsubsection{``So they can drop their guard a little bit.''}
\begin{figure}[ht]
  
  \includegraphics[width=0.9\columnwidth]{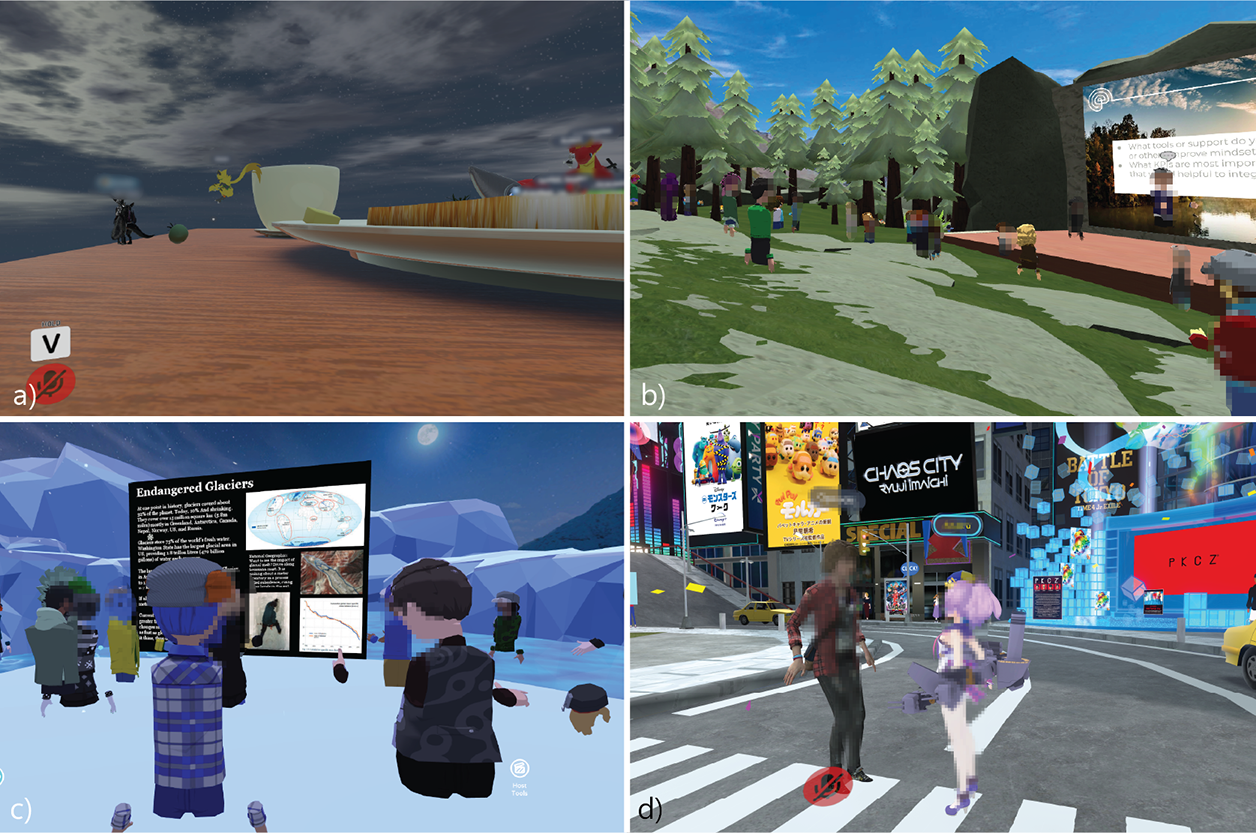}
  \caption{Examples of professional events observed in our research and featured scenes that were unconventional.}
  \Description{Professional events were hosted in virtual venues featured scenes that were unconventional.}
  \label{fig:differentvrspace}
\end{figure}

Event venues observed in our research showed large variations in the overall ambiance of their spatial design. Some were crafted to emulate the formality of real-world conferences, while others featured scenes that were rather unconventional (Figure~\ref{fig:differentvrspace}). For participants who hosted their own events in social VR, the freedom to decide the atmosphere of a virtual venue was highly valued. As commented by P12 and P13: 

\begin{quote}
    \textit{``My community is full of people doing animations and digital arts. Our field of work values creativity, innovation, and freedom. So, in some of the virtual exhibitions or seminars, I want the event to feel less tense. I have tried setting up really random worlds in VR, like putting in trees with bright purple fruits. When people come to these events, they may think, ‘oh, this place is cool!' They may find it easier to talk to others without worrying about behaving formally.''} [P12, Host, VRChat]
\end{quote}

\begin{quote}
    \textit{``At events made for professional purposes, people should still be able to DEW—Discover, Explore, and Wonder—and part of that involves play. It sometimes means doing things that you can't do in real life because you would be limited by physical and social rules. VR helps us create a space that’s playful and welcomes all sorts of explorations. People like it because they can be more true to themselves.''} [P13, Host, AltSpaceVR]
\end{quote}

\subsubsection{``What can be done or what cannot?''}
In contrast to the euphuism held by hosts, attendees who visited venues with unconventional spatial designs expressed mixed feelings. While many appreciated the freedom to \textit{“just be yourself in this hyperreal place,”} people noted that the freedom might come with a potential cost that matters to their career. 

\begin{quote}
    \textit{``If there were people in the crowd who were not internet native, didn’t grow up playing games, and were at this professional event, they might see my moments of being exploratory or relaxed as unprofessional. That person could,  in real life,  be a tech founder or someone that I’d like to connect. But they wouldn't take me seriously even though I might be a person they would actually talk to in real life.''} [P3, Attendee, AltSpaceVR]
\end{quote}

\begin{quote}
    \textit{``You don’t know what you’re able to do to or allowed to do. You see the setup and you feel you can actually do whatever you want in VR. But then you think this is a formal space to meet other real people. It’s hard to tell what you’re allowed to do even if you are clearly standing in an auditorium.''} [P2, Attendee, VRChat]
\end{quote}

\section{Discussion and Design Implication}
To recap, participants in our study intuitively drew upon their experiences from physical networking events when navigating events hosted in social VR. The bulk of incidents documented in our data confirm participants' awareness that the social implications of spatial information in these two worlds are not fully equivalent. However, few of them hold a precise understanding of the boundaries of spatial-as-social-cues between physical and virtual environments. Existing VR platforms fail to support this sensemaking. Below, we outline four directions for future VR systems to assist networking events from the perspective of spatial design and for both event hosts and attendees.

\subsection{Facilitating Spatial Understanding with Historical Information}
As illustrated in Section~\ref{resize}, the size of a virtual area is crucial for facilitating networking events. However, existing VR platforms reviewed in our work do not offer sufficient support regarding spatial understanding—determining the appropriate size is challenging for event hosts, and navigating through spaces of varying sizes is difficult for attendees. 

Hosts often rely on their anecdotal experience when deciding on the size of a virtual event. While they can take advantage of the flexibility of VR to adjust space size between event iterations, they lack the tools to effectively determine the appropriate size of a virtual space from the outset. For attendees, encountering virtual spaces with a mismatched number of others makes the area appear overly crowded or empty, leading to their quitting the event. 

To support the design of VR spaces and improve navigation through areas with varying perceived crowdedness, we suggest that future VR platforms provide more detailed information to both hosts, during the space design stage, and to attendees, during the events themselves.

In particular, when a host wants to design a new space for a networking event, a VR platform could provide historical attendance numbers of similar events, using aggregated data to approximate the number of attendees and thus suggest the proper size of a given virtual area. Such a system could reduce the reliance on event hosts' personal experience for space design, and could provide a better experience for attendees starting from the first venue.

Similarly, attendees could benefit from a VR platform that provides additional information about the event they are attending. For instance, future VR platforms might offer real-time participant distribution across different virtual areas, e.g., using a heatmap overlay on top of a minimap of the virtual space. Attendees could then use this information to calibrate their perceptions of crowdedness or emptiness in their current space, or follow the heatmap to move to areas with fewer or more people for networking. While the specific interface implementation can vary, we expect that such global information will provide attendees with much-needed spatial awareness when, for example, the first-person multi-sensory perception is just not sufficient, as described in Section~\ref{audio}.

\subsection{Optimizing Virtual Pathway Design for Social Networking} 
From Section~\ref{pathways}, we saw inspiring examples where event hosts create virtual pathways to aggregate, direct, and influence the foot traffic of event attendees, thereby creating opportunities for attendees to meet each other. Attendees also leverage such connection areas to better prepare for the transition between virtual areas. 

There is reason to believe that systematic guidance of the spatial layout design can greatly enhance networking opportunities at the target event. Unfortunately, all hosts in our study described the process of creating the virtual space as their intuitive practices. These practices varied much according to the tacit knowledge and experience held by each individual.

We argue that future VR platforms should provide users with space design recommendations based on the social networking opportunities they enable. Specifically, the VR design tool should focus not only on the visual rendering of a space but also on providing hosts with insights into the potential walk-through traffic that the design layout enables. For example, a future VR system might provide simulations of the expected attendee movement for a given space design and offer alternative space layout options that better align with the design goals specified by the host. While similar features have not been seen in commercial VR event design tools, simulating pedestrian flow has been available to architects when designing real physical buildings, streets, and city planning~\cite{simwalk, azhar2011building}, and can potentially be ported into VR design systems. 

Moreover, some of our findings reveal the tension between teleporting, which may be preferred by attendees, and the pathways designed by hosts. A future VR system may consider providing aggregate data on when and where attendees decide to teleport historically, constituting a different form of pedestrian flow. This information can facilitate the host's as well as the attendee's preconception regarding which pathways would be more frequently used at the event.

\subsection{Eliciting Spontaneous Interactions with the Implementation of Magnets} 
As we demonstrated in Section~\ref{magnet}, participants attending professional events in social VR displayed a greater sense of autonomy compared to those navigating social interactions via other digital mediums. Much of this autonomy is derived from the fact that people can actively reshape building blocks of the virtual venue. 

We have seen inspirational cases where hosts intentionally placed magnets in the virtual space to guide the attendee's attention towards one another; similarly, attendees created their own magnets as an effective means to initiate conversations with others. These practices have positive implications for networking at professional events, as they enhance the participant's self-control over potential opportunities for meaningful interactions. 

In line with this spirit, we suggest that future VR platforms should enable users to generate magnets and integrate them into a given VR space through diverse ways. Besides having the host place preset 3D artifacts ahead of the event, the system may allow attendees to bring in custom magnets designed by themselves. Such custom magnets can be 3D-scanned copies of an attendee's personal items, indicating the identity or image this person wishes to project. Alternatively, there can be 3D crafting or modeling functions built into the VR system (e.g., ~\cite{hui2024make, liu2022mix3d}), enabling an attendee's creation of their own magnets during the virtual events.

\subsection{Regulating Social Behaviors Through an Integrative Set of Guidance} 
With greater flexibility and autonomy afforded by VR, it can be challenging for everyone at the professional event to agree on a unified norm of behavior and adhere to it, as uncovered in Section~\ref{conventionality}. 

Several studies in social VR have examined such challenges and proposed mechanisms to explicitly regulate people's behaviors. For example, Schulenberg et al. suggested including computer-mediated consent mechanics at each of the social VR events, where attendees may specify the behaviors to which they consent~\cite{schulenberg2023we}. Blackwell et al. discussed the potential of implementing responsive regulation at social VR events. Specifically, upon a person's first time violating the social norm in VR, they may be informed to make self-correction before facing punishment or escalated sanctions~\cite{blackwell2019harassment}.

Reflecting on the relationship between prior work and ours, we believe the social protocol contextualized in space is essential but peripheral. When the spatial design of a VR event aligns with explicit behavior guidance (e.g., ~\cite{schulenberg2023we,blackwell2019harassment}), it can serve as an ambient clue reminding people of the appropriate behaviors to perform. That said, we cannot count on the spatial design alone to adequately guide everyone's behaviors in VR. 

One empirical question for future research to consider is how different spatial design strategies may consolidate or diminish people's understanding of the behavioral norms explicitly communicated through other mechanisms. To that end, our current work has identified a set of connections between VR spatial design and networking behaviors for follow-up studies to build upon.

\section{Limitations}
The findings from our research should be interpreted with certain limitations in mind. 
By using a qualitative research approach that included in-situ observations and in-depth interviews, we gained a nuanced understanding of our participants' experiences and behaviors in these virtual spaces. However, this method limited the number of participants we could involve in the study. While we aimed to study professional events across diverse social VR platforms and included participants taking different roles, our final sample may not fully capture the complete spectrum of situations that relate to our RQ. The networking structure among all attendees of a given event was not fully analyzed either. The presence of the researcher at each event might also inadvertently influence the participant's behavior, although it would otherwise be difficult to obtain the level of detail we sought while still maximizing the participant’s control of their disclosure. 

\section{Contribution and Conclusion}
In this paper, we contribute an in-depth qualitative insights into the interplay between spatial design and professional networking in a social VR context. 
Through in-situ observations and in-depth interviews with 13 participants across 20 professional networking events, we identified four sets of associations between the spatial design of a VR-based professional event and people's networking behaviors at the event. Our data evidenced how the size of a virtual area could influence a person's belief in their likelihood of encountering others during the professional event, how pathways between areas could guide people's planning of the next social interactions to perform, how strategically placed magnets in an area could enable self-initiated gatherings among strangers, and how the conventionality of each scene could help individuals assess the appropriateness of their behavior in it. These findings underscore the value and necessity of enhancing people's understanding of spatial-as-social clues in VR-based professional events by incorporating historical information from similar events, simulating the anticipated movements of others, enabling the convenient import of user-generated magnets, and integrating both ambient and explicit guidance for behavior regulation.

\begin{acks}
We thank all participants for their valuable time and insights. ChatGPT is used in this work solely to correct grammar errors.
\end{acks}

\bibliographystyle{ACM-Reference-Format}
\bibliography{citation}

\end{document}